\documentclass[onecolumn]{aastex62}
\usepackage[utf8]{inputenc}
\usepackage[T1]{fontenc}

\usepackage[version=4]{mhchem}
\usepackage[inline]{enumitem}
\usepackage{gensymb}
\usepackage{float}
\usepackage{fdsymbol}
\usepackage{amsmath}
\usepackage{xcolor}

\usepackage{xcolor}
\usepackage{xifthen}
\usepackage[normalem]{ulem}
\newcommand{\stkout}[1]{\ifmmode\text{\sout{\ensuremath{#1}}}\else\sout{#1}\fi}
\newcommand{\edited}[2]{\ifthenelse{\isempty{#1}}{\textcolor{red}{#2}}{\ifthenelse{\isempty{#2}}{\textcolor{gray}{\stkout{#1}}}{\textcolor{gray}{\stkout{#1}} \textcolor{red}{#2}}}}

\newcommand{\hms}[3]{#1\mathrm{h}#2\mathrm{m}#3\mathrm{s}}
\newcommand{\dms}[3]{#1\degr#2\arcmin#3\arcsec}

\received{May 31, 2019}
\revised{July 14, 2019}
\accepted{July 15, 2019}

\shorttitle{Vibrationally Excited \ce{CH3COOH} toward NGC 6334I}

\begin{document}
\title{ALMA Detection of Vibrationally Excited ($v\mathrm{_t} = 1,2$) Acetic Acid toward NGC 6334I}


\author{Ci Xue}
\affiliation{Department of Chemistry, University of Virginia, Charlottesville, VA 22904, U.S.A}

\author{Anthony J. Remijan}
\affiliation{National Radio Astronomy Observatory, Charlottesville, VA 22903, U.S.A}

\author{Crystal L. Brogan}
\affiliation{National Radio Astronomy Observatory, Charlottesville, VA 22903, U.S.A}

\author{Todd R. Hunter}
\affiliation{National Radio Astronomy Observatory, Charlottesville, VA 22903, U.S.A}

\author{Eric Herbst}
\affiliation{Department of Chemistry, University of Virginia, Charlottesville, VA 22904, U.S.A}
\affiliation{Department of Astronomy, University of Virginia, Charlottesville, VA 22904, U.S.A}

\author{Brett A. McGuire}
\altaffiliation{B.A.M. is a Hubble Fellow of the National Radio Astronomy\\ Observatory.}
\affiliation{National Radio Astronomy Observatory, Charlottesville, VA 22903, U.S.A}
\affiliation{Center for Astrophysics $\mid$ Harvard~\&~Smithsonian, Cambridge, MA 02138, USA}

\correspondingauthor{Ci Xue, Brett A. McGuire}
\email{cx5up@virginia.edu, bmcguire@nrao.edu}

\begin{abstract}
Vibrationally excited states of detected interstellar molecules have been shown to account for a large portion of unidentified spectral lines in observed interstellar spectra toward chemically rich sources. Here, we present the first interstellar detection of the first and second vibrationally excited torsional states of acetic acid ($v_\mathrm{t} = 1, 2$) toward the high-mass star-forming region NGC 6334I. The observations presented were taken with the Atacama Large Millimeter/submillimeter Array in bands 4, 6, and 7 covering a frequency range of 130 -- 352 GHz. By comparing a single excitation temperature model to the observations, the best-fit excitation temperature and column density are obtained to be 142(25) K and $1.12(7) \times 10^{17} \mathrm{cm^{-2}}$ respectively. Based on the intensity maps of the vibrationally excited \ce{CH3COOH} transitions, we found that the \ce{CH3COOH} emissions are compact and concentrated toward the MM1 and MM2 regions with a source size smaller than 2\arcsec. After locating the emission from different \ce{CH3COOH} transitions, which cover a large range of excitation energies, we are able to explain the variation of the \ce{CH3COOH} emission peak within the MM2 core by invoking continuum absorption or outflows.

\end{abstract}

\section{Introduction}
Although more than 200 individual molecular species have been detected in both galactic and extragalactic interstellar and circumstellar environments to date \citep{2018ApJS..239...17M}, a substantial number of unidentified spectral lines in the observed spectra toward a host of molecule-rich regions continue to plague observational astrochemistry. While many of these features are certainly due to as of yet unidentified new species, the carriers of these unidentified lines could also be isotopologues or vibrationally excited states of already known molecules \citep{2013ApJ...768...81D, 2014A&A...572A..44L, 2016A&A...595A..87M}. As such, studying the vibrationally excited states of known molecules is essential to characterizing the observed spectra and interstellar chemical inventories \citep{2015ApJ...803...97S}.

Acetic acid (\ce{CH3COOH}) is a large astronomical molecule (LAM) and of increasing interest to study from both an astrochemical and an astrobiological point of view \citep{2003ApJ...590..314R}. The first detection of interstellar \ce{CH3COOH} was toward Sgr B2(N) as reported by \citet{1997ApJ...480L..71M} and later confirmed by \citet{2002ApJ...576..264R}. To date, \ce{CH3COOH} has been detected in its vibrational ground state toward a variety of high- and low-mass star forming regions \citep{2003ApJ...593L..51C, 2003ApJ...590..314R, 2010ApJ...716..286S}. Since this initial detection, additional theoretical and experimental spectroscopic studies of the excited torsional states of \ce{CH3COOH} have provided the data needed to search for \ce{CH3COOH} in its first three torsional states ($v_\mathrm{t} = 0, 1,2$) up to 400 GHz \citep{2001JMoSp.205..286I, 2003JMoSp.220..170I, 2013JMoSp.290...31I}. Here, we present a search for the vibrationally excited rotational transitions ($v_\mathrm{t} = 1,2$) in an appealing target: NGC 6334I.
 
NGC 6334I is a massive star-forming core located at the northeastern end of the NGC 6334 molecular cloud \citep{2008A&A...481..169B} with a distance of ${\sim} 1.3\,\mathrm{kpc}$ \citep{2014ApJ...783..130R}. The millimeter continuum emission maps have spatially resolved NGC 6334I into nine distinct sources, with most of the hot core molecular line emission originating from the two brightest continuum sources, MM1 and MM2 \citep{2006ApJ...649..888H, 2016ApJ...832..187B}. MM1 is furthered characterized as a hot multi-core which comprises seven components and exhibits a high-velocity bipolar outflow \citep{2016ApJ...832..187B,2018ApJ...866...87B}. NGC 6334I is complex not only in physical structure but also in chemical composition \citep{2018A&A...615A..88B}. Rich spectral line emission originates from this region \citep{2000MNRAS.316..152M, 2006A&A...454L..41S}, as exemplified by interferometric line surveys with the Submillimeter Array (SMA) \citep{2012A&A...546A..87Z}, and the Atacama Large Millimeter/submillimeter Array (ALMA), the latter of which led to the first interstellar detection of methoxymethanol (\ce{CH3OCH2OH}) \citep{2017ApJ...851L..46M,2018ApJ...863L..35M}.

Recent observations of glycolaldehyde at ALMA Band 10 in this source showed bright emission lines from rotational states with upper state energies ($E_\mathrm{u}$) in the range of 530--631 K \citep{2018ApJ...863L..35M}. Combined with the high dust temperatures of MM1 \citep[up to 450~K;][]{2016ApJ...832..187B}, sufficient energy may be available to populate energetic low-lying vibrational states of LAMs that can then be detected through their rotational transitions within these excited states. Following the detection of \ce{CH3COOH} in the vibrational ground state toward NGC 6334I (El-Abd et al. 2019, under revision), NGC 6334I is clearly an ideal region to search for emission features of the vibrationally excited transitions of \ce{CH3COOH}, particularly because of its relatively narrow lines.

In this paper, we report the first interstellar detection of vibrationally excited \ce{CH3COOH} ($ v_\mathrm{t} = 1,2$) toward NGC 6334I with ALMA in Bands 4, 6 and 7. The observations are described in Section \ref{sec:obs}. The process of line identification and spatial imaging is presented in Section \ref{sec:dataanlysis}. In Section \ref{sec:diss}, the spatial distribution is discussed with respect to the complicated structure of the dust emission cores, MM1 and MM2. The results are summarized in Section \ref{sec:sum}.

\section{OBSERVATIONS} \label{sec:obs}
The detection of the first and second vibrationally excited states of \ce{CH3COOH} was achieved with three data sets involving ALMA Bands 4, 6, and 7 ranging from 130 to 352 GHz. The Band 6 data are first presented in this paper, while a detailed description of the Band 4 and 7 data have been presented elsewhere \citep{2018ApJ...863L..35M, 2017ApJ...837L..29H, 2017ApJ...851L..46M}. The Band 6 observation was performed on 2017 December 28 (project code 2017.1.00370.S). A synthesized beam size of $\mathrm{0.18 \arcsec \times 0.15 \arcsec}$ was achieved with a baseline range from 15 m to 2.5 km. The salient observational parameters for the Band 6 data are summarized in Table \ref{obs-par}. The Band 4 data (project code 2017.1.00661.S) were taken in Cycle 5 with a spectral resolution of 0.488 MHz and an rms noise of $\mathrm{\sim 0.8\ mJy\,beam^{-1}}$ \citep{2018ApJ...863L..35M}. The Band 7 data (project code 2015.A.00022.T) were taken in Cycle 3 with a spectral resolution of 0.977 MHz and an rms noise of $\sim 2.0$ mJy\,beam$^{-1}$ at 1.1 mm and $\mathrm{3.3\ mJy\,beam^{-1}}$ at 0.87 mm \citep{2017ApJ...837L..29H, 2017ApJ...851L..46M}.

The data have been analyzed with the Common Astronomy Software Applications (CASA) package \citep{2007ASPC..376..127M}. The relatively line-free continuum channels were selected with the techniques described in \citet{2018ApJ...866...87B} and used to subtract the continuum emission in the UV-plane. The Briggs weighting scheme and a Robust parameter of 0.5 for the Band 4 and 6 data and 0.0 for the Band 7 data were used in the \texttt{TCLEAN} task of CASA for imaging, which resulted in elliptical beam shapes. The images were smoothed to a 0.26$\arcsec$ uniform, circular beam size from the image plane for consistency among the 3 datasets. The spectra presented in this paper were extracted toward MM1 from a single pixel centered at $\alpha_\text{J2000}=\hms{17}{20}{53.374}$, $\delta_\text{J2000}=\dms{-35}{46}{58.34}$.

\begin{deluxetable}{lc}
    \tablewidth{0.9pt}
    \tablecaption{Observational Parameters for the Band 6 ALMA Data \label{obs-par}}
    \tablehead{
        \colhead{Parameter}                 &\colhead{Band 6 (259 GHz)}}
    \startdata
        Observation Date                    &28th Dec 2017              \\
        Cycle                               &5                          \\
        Project Code                        &2017.1.00370.S             \\
        Configuration                       &C43-5/C43-6                      \\
        Time on Source (minutes)            &48                         \\
        Phase Center (J2000 R.A., decl)     &17:20:53.35, -35:47:01.5   \\
        SPW Center Frequency (GHz)               &251.4, 251.8, 266.3, 270.7 \\
        SPW Bandwidth (MHz)                   &468.75                     \\
        HPBW Primary Beam ($\arcmin$)       &0.37                       \\
        Gain Calibrator                     &J1713-3418                 \\
        Bandpass and Flux Calibrator        &J1517-2422                 \\
        Phase Calibrator                    &J1733-3722                 \\
        Spectral Resolution (MHz)            &0.244                      \\ 
        Ang. res \tablenotemark{a}($\arcsec \times \arcsec$)
                                             &0.18 $\times$ 0.15        \\
        rms per Channel ($\mathrm{mJy\,beam^{-1}}$)& 4.0 \\
   \enddata
    \tablecomments{
    \tablenotetext{a}{The angular resolution is achieved with a Robust weighting parameter of 0.5. The images are further smoothed to $\mathrm{0.26 \arcsec \times 0.26 \arcsec}$ for consistency between all observations.}
    }
\end{deluxetable}

\section{ANALYSIS AND RESULTS}\label{sec:dataanlysis}
\subsection{Line Identification} \label{sec:transitions}
The dense molecular spectra of NGC 6334I as described by \citet{2012A&A...546A..87Z} indicate a high degree of spectral contamination, i.e., a substantial proportion of spectral lines are overlapped and blended with adjacent lines of either the same or another species. To positively identify a vibrationally excited \ce{CH3COOH} line from this highly contaminated spectra, we adopt the five criteria for line identification proposed in \citet{2005ApJ...619..914S} (hereafter, the Snyder criteria): (1) accurate rest frequencies, (2) take into account beam dilution effect, (3) frequency agreement, (4) intensity agreement, and (5) presence of transitions with detectable intensity. Because the processes for analyzing following the Snyder criteria in this study are similar with those in \citet{2019ApJ...871..112X}, we will highlight the main points here, and refer the reader to Section 3.1 in \citet{2019ApJ...871..112X} for more detail.

The emitting area of most LAMs toward NGC 6334I is over an extent of $\sim 5 \arcsec$ \citep{2017ApJ...851L..46M}, while that of glycolaldehyde (\ce{CH2OHCHO}), a structural isomer of \ce{CH3COOH}, has a source size $\sim 2.5\arcsec\ \times\ 0.8\arcsec$ \citep{2018ApJ...863L..35M}. Accordingly, it is reasonable to assume that the spatial distribution of the \ce{CH3COOH} emission is more extended than the angular resolution used in this study ($0.26 \arcsec$), and therefore, the beam dilution effect is negligible. On the other hand, it is very unlikely that the source size of \ce{CH3COOH} would exceed the maximum recoverable scale of the ALMA observations presented in this study ($\sim 3 \arcsec$). If there is a large-scale emission of \ce{CH3COOH}, it should be observable with single dishes but \ce{CH3COOH} has never been detected toward NGC 6334I with single-dish observations. Moreover, the \ce{CH3COOH} emission display a compact morphology in many other hot molecular cores \citep{2003ApJ...590..314R}, which further support our assumptions.

\citet{2018ApJ...863L..35M} showed that a single-excitation temperature can well describe the spectra of most LAMs across the full range of ALMA wavelengths (Bands 3-10) toward the MM1 region, in which $n_\text{\ce{H2}} > 10^9~\mathrm{cm^{-3}}$ \citep{2016ApJ...832..187B}. After estimating the critical density for the \ce{CH3COOH} transitions in the vibrationally excited states, we found that $n_\mathrm{crit}$ of the $v_\mathrm{t} = 1$ and $v_\mathrm{t} = 2$ transitions are at most $\sim 10^5~\mathrm{cm^{-3}}$. The upper limit of $n_\mathrm{crit}$ is given by a lower limit of the collisional cross section, which is ${\sim} 10^{-15}\,\mathrm{cm^2}$, the cross section of \ce{H2}. It is noted that the cross section of \ce{HCOOCH3}, a molecule that has similar complexity with \ce{CH3COOH}, is significantly larger than that of \ce{H2} \citep{2014ApJ...783...72F}. Hence, while being conservative, the upper limit of $n_\mathrm{crit}$ estimated here is well-grounded. As $n_\text{\ce{H2}} \gg n_\mathrm{crit}$, it is reasonable to assume that these \ce{CH3COOH} transitions are thermalized toward MM1. 

Following the convention of \citet{1991ApJS...76..617T}, we simulated the spectrum of \ce{CH3COOH} in each state with the assumption that the transitions of \ce{CH3COOH} are thermalized, as in \citet{2018ApJ...863L..35M}. Constant background continuum temperature ($T_\mathrm{bg}$) is considered in each frequency range covered by the datasets. In particular, based on brightness temperature measurements, $T_\mathrm{bg}$ is estimated to be $11.5~\mathrm{K}$, $27.4~\mathrm{K}$, $33.9~\mathrm{K}$ and $43.1~\mathrm{K}$ for the Band 4, Band 6, Band 7 1.1 mm and Band 7 0.87 mm datasets respectively. The spectroscopic data of the vibrationally excited states in the 130-352 GHz frequency range are taken from \citet{2013JMoSp.290...31I}. The torsional vibrational states of \ce{CH3COOH} have the following excitation energies: the $v_\mathrm{t} = 1$ state lies at around $74~\mathrm{cm^{-1}}$ (or $107~\mathrm{K}$) while the $v_\mathrm{t} = 2$ state lies at $127~\mathrm{cm^{-1}}$ (or $184~\mathrm{K}$) above the ground state \citep{2001JMoSp.205..286I}. Within the observed spectral range, the criterion of accurate rest frequencies is fulfilled by the small uncertainties associated with the transitions, which are smaller than $0.2~\mathrm{km\,s^{-1}}$, compared with the observed linewidth (${\sim} 3~\mathrm{km\,s^{-1}}$). The torsional-rotational partition function $Q_\mathrm{rv}$ is approximated by fitting the total internal partition function as a function of temperature ($T$) using a sixth order polynomial \citep{2013JMoSp.290...31I}:
\begin{equation}
    \begin{split}
        Q_\mathrm{rv}(T) = & -8.88 \times 10^{-11}\,T^6 + 1.03 \times 10^{-7}\,T^5 \\
                           & - 5.02 \times 10^{-5}\,T^4 + 1.34 \times 10^{-2}\,T^3 \\
                           & + 7.83 \times 10^{-1}\,T^2 + 49.0 \,T - 139 .
    \end{split}
\end{equation}

By assuming $T_\mathrm{ex}$ to be 135 K \citep{2018ApJ...863L..35M}, 40 spectral emission features for $v_\mathrm{t} = 1$ and 17 spectral emission features for $v_\mathrm{t} = 2$ are expected to be over the detection threshold, i.e, with $S/N > 3$, in this study. Only a few of these line features are individual transitions, while the others are an ensemble of multiple transitions. We find all these emission lines to be present and no lines to be missing in the observed spectra if we take into account the contaminated lines, as shown in Appendix \ref{sec:a-spect}.

\citet{2019ApJ...871..112X} describe a way to quantitatively measure the agreement between the modeled spectra and the observations by, first, selecting an appropriate spectral range for comparing line profiles from the observation and simulation and, second, computing the $P$ (Product) and $D$ (Difference) factors \citep[see equations (1)-(2) of][]{2019ApJ...871..112X}. The $P$ factor is based on the product of the observed and modeled line profiles to characterize how close the observed line is located to the expected rest frequency. The $D$ factor is based on the difference of the integrated intensities between the spectra, which compares the modeled line intensities to the astronomically observed values. Instead of setting an absolute threshold for the $P$ and $D$ factors, it is more relevant to compare the derived factors among each emission feature; overall, features with higher $P$ and $D$ factors are more likely to be ``clean''. Therefore, the two factors serve as tools to assist and accelerate the line identification process by eliminating features that are buried under or strongly contaminated by other emission features. This greatly reduces the number of emission features to be visually examined and provides a feasible and crucial strategy for line identification over large bandwidth. Due to the success of uniquely identifying lines in \citet{2019ApJ...871..112X}, the same approach is adopted here.

\begin{figure*}[ht]
    \centering
    \epsscale{1.2}
    \plotone{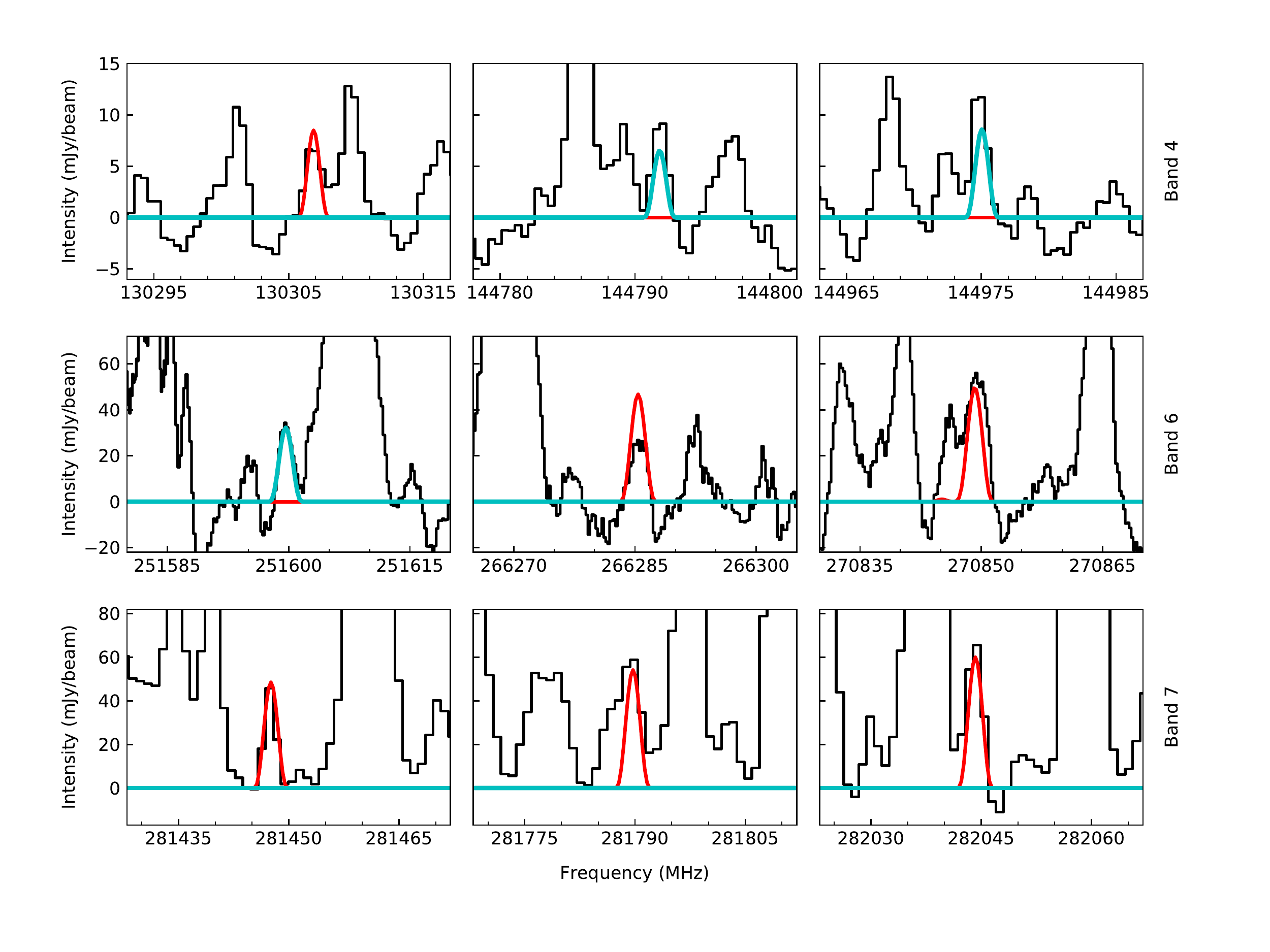}
    \caption{Band 4, 6, 7 spectra of NGC 6334I extracted toward the MM1 region are shown in black. The best-fit simulated spectra are overlaid in red and cyan for $v_\mathrm{t} = 1$ and $v_\mathrm{t} = 2$ respectively. The best-fit parameters are $T_\mathrm{ex} = 142~\mathrm{K}$, $N_\mathrm{T} = 1.12 \times 10^{17}~\mathrm{cm^{-2}}$, and $\Delta V = 2.72~\mathrm{km\,s^{-1}}$. The horizontal axes are the rest frequency with respect to a radial velocity of $-6.9~\mathrm{km\,s^{-1}}$. Only the 9 identified emission features are shown in this figure. Refer to Appendix \ref{sec:a-spect} for the other transitions with detectable expected intensities.
    \label{v1_and_2-spec}}
\end{figure*}

We identify a total of six $v_\mathrm{t} = 1$ features and three $v_\mathrm{t} = 2$ features, as shown in Figure \ref{v1_and_2-spec}. All of these \ce{CH3COOH} features are a collection of four transitions. The observational results of the clean emission features are summarized in Table \ref{v-table}, while the detailed spectroscopic parameters of the pertaining transitions are listed in Appendix \ref{sec:transitions_detail}. Note that the 281789 MHz line complex has a broader line width than the others, which may be due to the presence of a weak blended emission feature located near the wing of the 281789 MHz feature, as will be discussed in Section \ref{sec:map}.

The excitation temperature ($T_\mathrm{ex}$) and the column density ($N_\mathrm{T}$) are determined by fitting a single excitation temperature model to the observed data. The observed results of the clean emission features are used to constrain the line width ($\Delta V$) and source velocity ($V_\mathrm{lsr}$), which are in good agreement among all the clean features. Therefore, $\Delta V$ is fixed to the median of the FWHM of the clean line complexes and $V_\mathrm{lsr}$ is fixed to the median value. $\chi^2$ fitting is performed to statistically optimize the model and determine $T_\mathrm{ex}$ and $N_\mathrm{T}$. Rather than considering all of the detectable emission features, the fitting procedures are performed with the clean features only. The process is also described in \citet[see section 3.3]{2019ApJ...871..112X}.

The $\Delta V$ is fixed to $2.72~\mathrm{km\,s^{-1}}$ and the $V_\mathrm{lsr}$ is fixed to $-6.90~\mathrm{km\,s^{-1}}$. A $T_\mathrm{ex}$ of 142(25) K and an $N_\mathrm{T}$ of $1.12(7) \times 10^{17}~\mathrm{cm^{-2}}$ minimize the $\chi^2$ value and best fit the observations. The values of $\Delta V$ and $V_\mathrm{lsr}$ characterized here are consistent with the values obtained from the spectrum of glycolaldehyde emission, $\Delta V = 3.2~\mathrm{km\,s^{-1}}$ and $V_\mathrm{lsr} = -7~\mathrm{km\,s^{-1}}$ \citep{2018ApJ...863L..35M}. The $N_\mathrm{T}$ and $T_\mathrm{ex}$ also correlate fairly well with El-Abd et al.\ (2019, under revision), who measure an $N_\mathrm{T}$ of $1.24 \times 10^{17}~\mathrm{cm^{-2}}$ and a $T_\mathrm{ex}$ of 135 K derived from the \ce{CH3COOH} transitions in the ground state toward the MM1 region. The uncertainties quoted here are the standard deviations ($1 \sigma$) of the best-fit model, which include the fitting error derived from the inverse of the Hessian matrix and the error propagation from the rms noise of the observed spectra. We want to emphasize that the uncertainties listed here might be underestimated as we have not incorporated the systematic uncertainties introduced by continuum subtraction.

\startlongtable
\begin{deluxetable*}{cccccccc}
    \tablewidth{0pt}
    \tablecaption{Ensemble of Spectral Line Transitions of Vibrationally Excited \ce{CH3COOH} \label{v-table}}
    \tablehead{
        \colhead{Emission Features} &\colhead{$v_\mathrm{t}$}    &\colhead{$I_\mathrm{peak}$}         &\colhead{FWHM}         &\colhead{$V_\mathrm{lsr}$}  &\colhead{P Factor} &\colhead{D Factor} &\colhead{Channel rms}\\
        \colhead{(MHz)} &~  &\colhead{($\mathrm{mJy\,beam^{-1}}$)}   &\colhead{($\mathrm{km\,s^{-1}}$)}  &\colhead{($\mathrm{km\,s^{-1}}$)} &~  &~  &\colhead{($\mathrm{mJy\,beam^{-1}}$)}
        }
    \startdata
	    130306  &1  &6.605  &3.33   &-6.90  &98.7   &93.5   &0.8\\
	    144791  &2  &9.134  &2.59   &-6.96  &99.3   &58.1   &0.8\\
	    144975  &2  &11.697 &2.72   &-6.48  &95.6   &61.1   &0.8\\
	    251599  &2  &34.337 &2.36   &-6.95  &99.4   &90.7   &4.0\\
        266285  &1  &26.868 &2.27   &-6.98  &98.9   &62.8   &4.0\\
        270849  &1  &56.095 &3.07   &-7.10  &98.6   &68.4   &4.0\\
        281447  &1  &45.838 &2.00   &-6.76  &98.3   &88.1   &2.0\\
        281789\tablenotemark{a}  &1  &58.785 &4.05   &-6.43  &97.7   &68.1   &2.0\\
        282044\tablenotemark{b} &1  &65.64  &2.91   &-6.64   &98.3   &77.0   &2.0\\
    \enddata
    \tablecomments{Observed and fitted parameters of the clean emission features pertaining to vibrationally excited \ce{CH3COOH} ($v_\mathrm{t} = 1$ and $v_\mathrm{t} = 2$). The approximate frequencies listed here are used to represent the emission features, of which the accurate rest frequencies are presented in Appendix \ref{sec:transitions_detail}. FWHM and $V_\mathrm{lsr}$ were obtained by the Gaussian fitting function. P and D Factors were calculated based on the observed spectra toward the MM1 region and the simulated spectra with a $T_\mathrm{ex}$ of 142 K. \tablenotetext{a}{The 281789 MHz feature suffers from spatial contamination due to the weak blended emission feature at the line wing.} \tablenotetext{b}{The 282044 MHz feature suffers from spatial contamination due to the adjacent line.}}
\end{deluxetable*}

\subsection{Spatial Distribution}\label{sec:map}
We can only be confident in the spatial extent and morphology for those spectral regions where the extracted spectra fulfill the Snyder criteria. Thus, there remains the possibility of a problem of spatial contamination toward other regions of the same source, i.e., contamination from other molecules may be present in the regions where the spectra are not examined. Nonetheless, the chance of the maps being contaminated by molecules with a similar spatial distribution pattern is unlikely. If a similar morphology is displayed among most of the transitions of a given molecule, odds are that it represents the intrinsic spatial distribution of that molecule. Therefore, the spatial distribution coherence can serve as an additional criterion for securing the morphology determination \citep{2015A&A...576A.129B}.

\begin{figure*}[ht]
    \centering
    \epsscale{1.2}
    \plotone{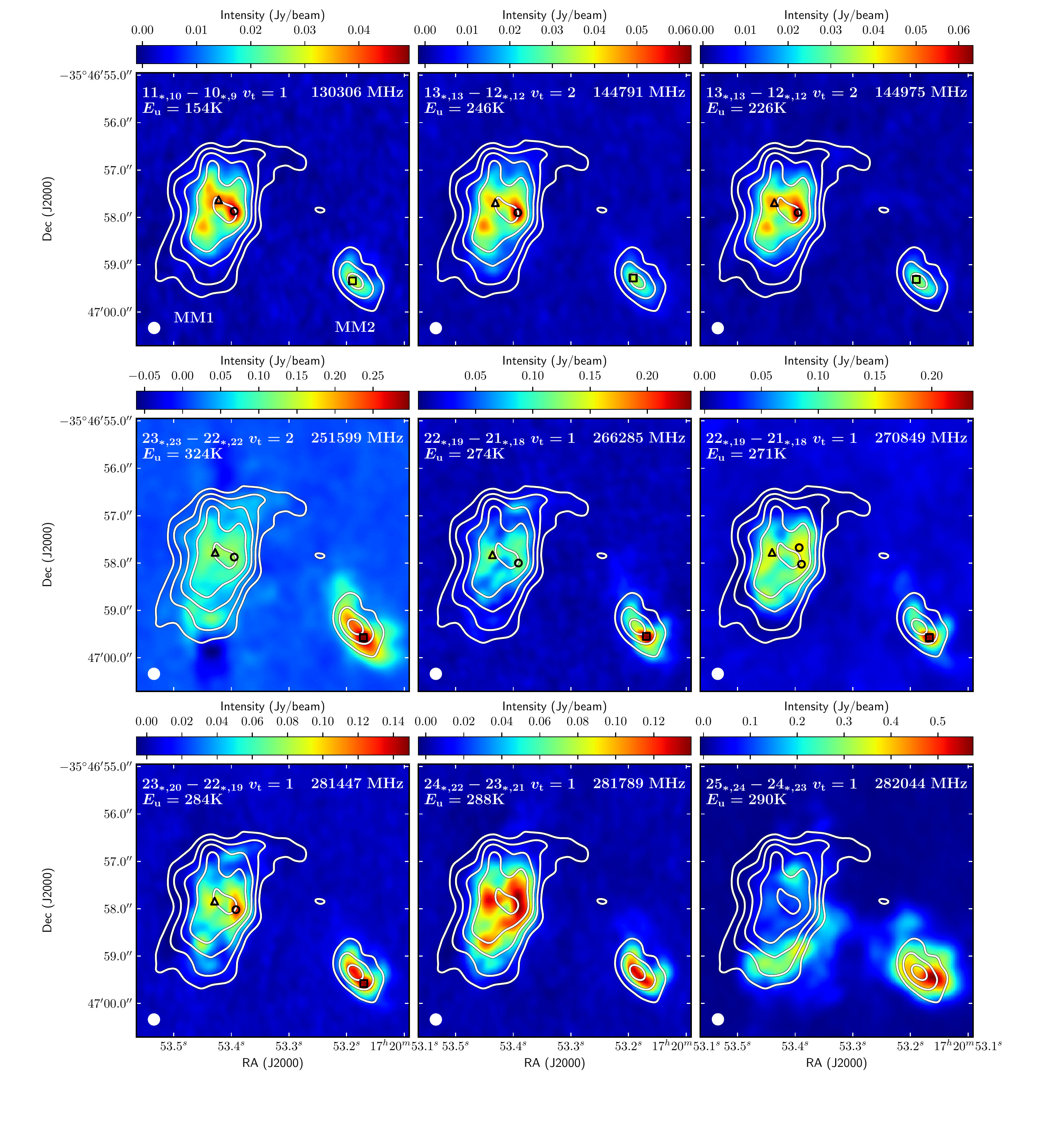}
    \caption{The peak intensity maps over the velocity range of $-3$~to~$-9 \mathrm{km~s^{-1}}$ of the clean lines of vibrationally excited \ce{CH3COOH} overlaid with the 287 GHz continuum emission in contours. The resolved quantum numbers represented by asterisks in the legends are listed in Appendix \ref{sec:transitions_detail}. The contour levels correspond to 46, 92, 184, 368 and 736 $\mathrm{mJy\,beam^{-1}}$. The images are smoothed with uniform circular beam size of $0.26 \arcsec \times 0.26 \arcsec$, as shown in the bottom left corner. The open circle, triangle, and square markers denote the \ce{CH3COOH} molecular emission peaks. The inconsistency seen in the morphology of the 281789 MHz and 282044 MHz maps compared with the others suggest the two maps suffer from spatial contamination from other molecular emission.
    \label{v1_and_2-mom8}}
\end{figure*}

The continuum emission features at Bands 4, 6, and 7 have different flux densities but display consistent morphology toward MM1 and MM2. In this study, we use the continuum emission map at 287 GHz from the Band 7 observations as representative, which is shown in contours in Figure \ref{v1_and_2-mom8}. All the 9 peak intensity maps of vibrationally excited \ce{CH3COOH} within the $V_\mathrm{lsr}$ range of $-3$ to $-9~\mathrm{km\,s^{-1}}$ are presented in color. However, the right two maps in the bottom row are discarded from the following analysis because of spatial inconsistencies and possible contamination as described below. The broad line width and relatively extended distribution of the 281789 MHz feature suggest the existence of contamination, which might be attributed to the propanal 26(7,19)-25(7,18) transition at 281787 MHz. The 282044 MHz map is spatially contaminated by the emission of the nearby line complex peaked at 282039 MHz (assuming a $V_\mathrm{lsr}$ at $-6.8 \mathrm{km s^{-1}}$), which is the superposition of \ce{S^{18}O}, \ce{c-HCCCH} and \ce{SO2} lines. Apart from the two contaminated maps, the seven remaining peak intensity maps of vibrationally excited \ce{CH3COOH} present a coherent distribution with each other over a range of energy levels. 

As is clear from the seven remaining maps, the emission from \ce{CH3COOH} is mainly distributed over both MM1 and MM2. The reported coordinates of MM1 and MM2 are ($\alpha_\text{J2000}=\hms{17}{20}{53.412}$, $\delta_\text{J2000}=\dms{-35}{46}{57.90}$) and ($\alpha_\text{J2000}=\hms{17}{20}{53.185}$, $\delta_\text{J2000}=\dms{-35}{46}{59.34}$), respectively. The source size of vibrationally excited \ce{CH3COOH}, estimated at half peak flux density, is resolved to be $\sim 2 \arcsec \times 1.2\arcsec$ toward MM1 and $\sim 1 \arcsec \times 0.4\arcsec$ toward MM2. Although its compact distribution is consistent with the continuum emission on a larger scale, they do not exactly coincide. There is an offset between the molecular emission and 287 GHz continuum emission peaks with a closest distance $< 0.36 \arcsec$. We resolved three prominent \ce{CH3COOH} emission peaks throughout all the clean line complexes. The AcA1 and AcA2 emission peaks, denoted by an open circle and triangle markers in Figures \ref{v1_and_2-mom8} and \ref{peak_position}, are within MM1, whereas the AcA3 emission peak marked with an open square is located toward MM2. The observed parameters, including the coordinates, intensity, and the distance between the molecular and continuum emission, of all the \ce{CH3COOH} emission features are listed in Table \ref{position-table}. The individual \ce{CH3COOH} peak positions can vary by at most $0.3\arcsec\/$ from one transition to another. 

\startlongtable
\begin{deluxetable*}{llccccc}
    \tablewidth{0pt}
    \tablecaption{Vibrationally Excited \ce{CH3COOH} Emission Peaks \label{position-table}}
    \tablehead{
        \\
       \colhead{Emission Features}  &\colhead{Peaks}   &\multicolumn{2}{c}{Coordinate} &\colhead{Dist}\tablenotemark{a} &\colhead{Intensity}    &\colhead{Markers}\tablenotemark{b}\\
        \colhead{(MHz)} &~  &\colhead{$17^h20^m$}   &\colhead{$-35 \degree 46\arcmin$}  &\colhead{($\arcsec$)}  &\colhead{($\mathrm{Jy\,beam^{-1}}$)}   &~       
        }
    \startdata
        130306&	AcA1& 	53.395s &57.870	&0.21	&0.0493	&$\medcircle$\\
              & AcA2&	53.422s &57.630	&0.30	&0.0405 &$\medtriangleup$\\
              & AcA3&	53.190s &59.340	&0.06	&0.0303	&$\medsquare$\\
        144791&	AcA1&	53.392s &57.900 &0.24	&0.0628 &$\medcircle$\\
              &	AcA2&	53.431s &57.690 &0.31	&0.0455 &$\medtriangleup$\\
              &	AcA3&	53.192s &59.280	&0.10	&0.0371	&$\medsquare$\\
        144975&	AcA1&	53.395s &57.900	&0.21	&0.0633 &$\medcircle$\\
              &	AcA2&	53.436s &57.690	&0.36	&0.0490	&$\medtriangleup$\\
              &	AcA3&	53.190s &59.310	&0.07	&0.0353	&$\medsquare$\\
        251599&	AcA1&	53.395s &57.875	&0.21	&0.1374 &$\medcircle$\\
              &	AcA2&	53.428s &57.775	&0.23	&0.1382 &$\medtriangleup$\\
              &	AcA3&	53.171s &59.575	&0.29	&0.2974 &$\medsquare$\\
        266285&	AcA1&	53.391s &58.000	&0.27	&0.1255 &$\medcircle$\\
              &	AcA2&	53.436s &57.825	&0.30	&0.1452 &$\medtriangleup$\\
              &	AcA3&	53.169s &59.550	&0.29	&0.2382 &$\medsquare$\\
        270849\tablenotemark{c}& AcA1&	53.389s &58.025	&0.31	&0.1545 &$\medcircle$\\
              &	AcA1&	53.393s &57.675	&0.32	&0.1584 &$\medcircle$\\
              &	AcA2&	53.440s &57.775	&0.36	&0.1513 &$\medtriangleup$\\
              &	AcA3&	53.167s &59.575	&0.32	&0.2364 &$\medsquare$\\
        281447&	AcA1&	53.392s	&58.020	&0.27	&0.1298 &$\medcircle$\\
              &	AcA2&	53.429s &57.840	&0.22	&0.0997 &$\medtriangleup$\\
              &	AcA3&	53.170s	&59.580	&0.30	&0.1490 &$\medsquare$\\
    \enddata
    \tablecomments{\tablenotetext{a}{The distance to the nearest 287 GHz continuum emission peaks, $\alpha_\text{J2000}=\hms{17}{20}{53.412}$, $\delta_\text{J2000}=\dms{-35}{46}{57.90}$ at MM1 and $\alpha_\text{J2000}=\hms{17}{20}{53.185}$, $\delta_\text{J2000}=\dms{-35}{46}{59.34}$ at MM2.}
    \tablenotetext{b}{The open circle, triangle, and square markers refer to the markers used in Figure \ref{v1_and_2-mom8} and Figure \ref{peak_position}.}
    \tablenotetext{c}{The 270849 MHz image has two emission peaks close to the AcA1 position.}
    }
\end{deluxetable*}

\section{DISCUSSION}\label{sec:diss}
\begin{figure}[ht]
    \centering
    \epsscale{0.7}
    \plotone{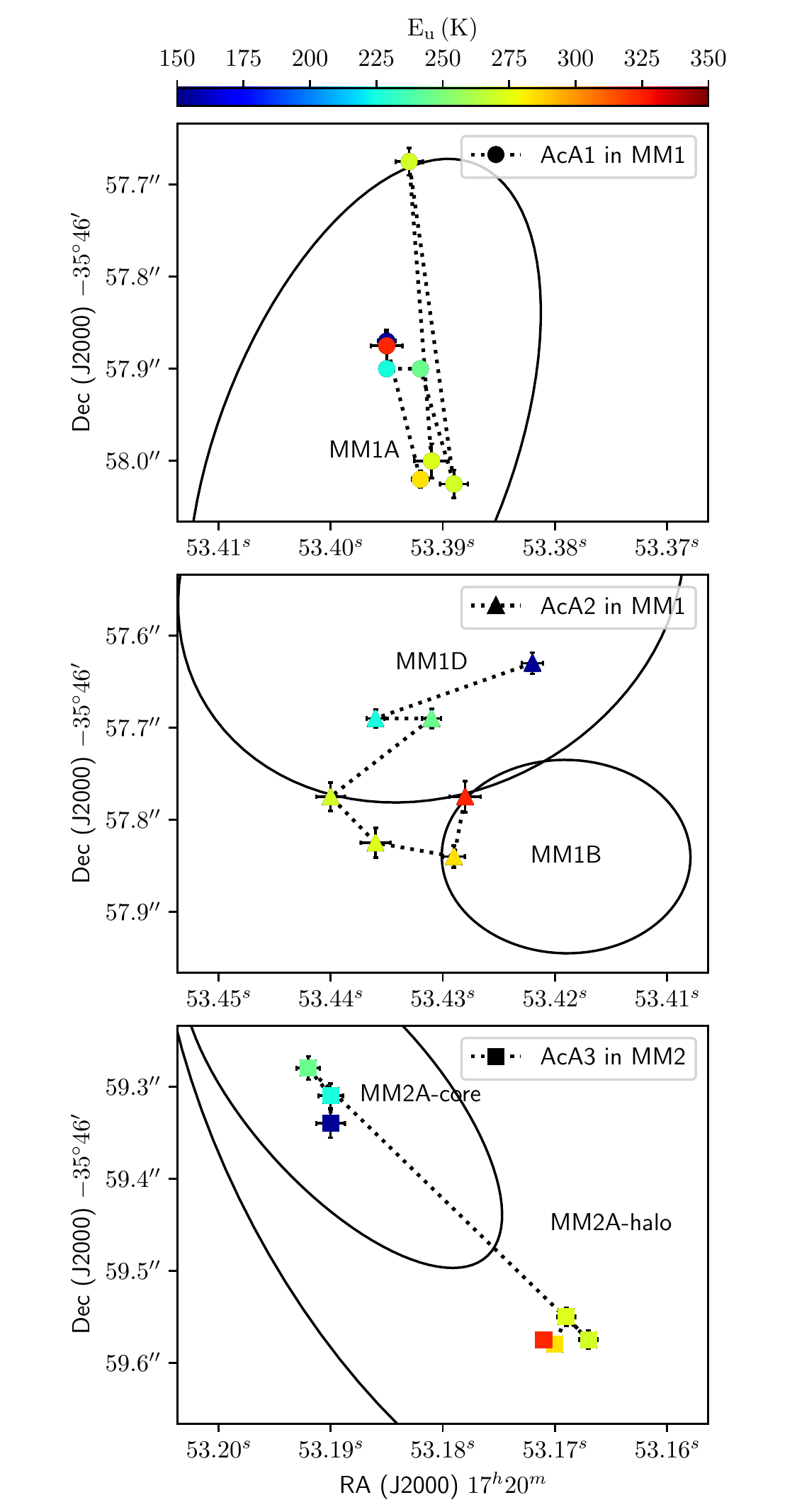}
    \caption{The peak emission positions for each of the seven spatially consistent transitions and the associated continuum emission cores. The peak emission positions close to AcA1, AcA2, and AcA3 are marked by circles, triangles, and squares, respectively. The colors of the markers represent the upper state energy of each transition. The error bars are the $5\sigma$ position uncertainties estimated by equation (1) in \citet{1988ApJ...330..809R}. The dotted lines connect the markers in ascending $E_\mathrm{u}$. The ellipses mark the extent of the dust cores MM1A, MM1B, MM1D and MM2A-core and -halo taken from \citet{2016ApJ...832..187B}. Note that the 270849 MHz map has two emission peaks close to the AcA1 position.
    \label{peak_position}}
\end{figure}

The multiple hot cores revealed toward NGC 6334I themselves exhibit complex structures and heterogeneous temperature distributions \citep{2016ApJ...832..187B}. Considering that the compact \ce{CH3COOH} emission is concentrated toward the MM1 and MM2 regions, we will discuss below the locations of the \ce{CH3COOH} emission peaks with regard to the sub-components of the two hot cores, MM1 and MM2. 

Seven dust clumps were resolved toward the hot core MM1, namely MM1A--G, with associated dust temperatures ranging from 116 to 450 K \citep{2016ApJ...832..187B}. As illustrated in Figure \ref{peak_position}, all the AcA1 emission peaks are associated with the densest core MM1A, which has $n_\mathrm{\ce{H2}} \sim 7.2 \times 10^9~\mathrm{cm^{-3}}$. The individual positions of the AcA1 peaks among different rotational states are scattered in a region the size of a synthetic beam. While the AcA2 emission peaks also show a dispersed distribution, they lie between the MM1B and MM1D compact components, which are the hottest dust sources of MM1 with $T_\mathrm{dust}$ at 442 K for MM1B and 305 K for MM1D \citep{2016ApJ...832..187B}. A distribution that lies in between cores is an atypical phenomenon of LAMs. Another example of a molecule with this kind of distribution is acetone ((CH$_3$)$_2$CO), which is located between the hot cores and compact regions in Orion KL \citep{2013A&A...554A..78P}. While the different peak positions versus $E_{\rm u}$ may indicate additional unresolved structures that are smaller than the beam, the complexity of the MM1 core makes the explanation and generalization of the positions of the \ce{CH3COOH} emission peaks difficult. 

In contrast, a core-halo structure was suggested for MM2, within which the dust temperature of the core was estimated to be $\sim 152$~K \citep{2016ApJ...832..187B}. Different from AcA1 and AcA2, which show dispersed distributions, the AcA3 peaks toward the MM2 cluster into two groups depending on their energy levels, which are separated by $0.3 \arcsec$. In particular, whereas the AcA3 emission peaks pertaining to the lower excited states of \ce{CH3COOH} are located toward the core region, the higher energy transitions peaks are located in the MM2 halo toward the south-west. This offset could be explained in part by the dust opacity. We have noticed that there is deep contaminating absorption from continuum emission at the high frequency bands (i.e., Bands 6 and 7). With the absorption toward the continuum peak becoming strong, the observable molecular emission toward the continuum peak would decrease and, consequently, the molecular emission peaks would deviate from the continuum peak. The effect of the strong contaminating absorption from continuum emission on our study is that the AcA3 emission peaks of the Band 6 and 7 transitions, with $\nu > 250$ GHz, would be offset from the continuum peak, whereas the AcA3 peaks of the Band 4 transitions, where there is almost no continuum absorption, are located toward the core region. Additional support for this explanation is provided by the existence of another \ce{CH3COOH} emission peak located to the southeast of the AcA2 peaks and close to the MM1C continuum core, which is only seen in the three Band 4 images. The disappearance of the emission peak from the higher frequency images could be explained well by the dust opacity in a similar fashion to the case of the AcA3 position shift.

Only a few studies have focused on spatially mapping vibrationally excited molecules and comparing the morphological differentiation of a molecule in different energy levels. The \ce{HCOOCH3} observation toward Orion KL made with ALMA indicated that the distributions of different vibrationally excited \ce{HCOOCH3} states resemble each other with a consistency of peak positions \citep{2015ApJ...803...97S}. In contrast, the work by \citet{2017ApJ...837...49P} with the same set of data showed a trend of the displacements among the \ce{HC3N} emission peaks. The \ce{HC3N} observation revealed that the emission peaks of its vibrationally excited lines are displaced from the south of the hot core region to the northeast as the $E_\mathrm{u}$ increases, which is coincident with the major axis of the \ce{SiO} outflow \citep{2017ApJ...837...49P}. Accordingly, the variation of the positions of the AcA3 peaks might also hint at an outflow structure proceeding in the southwest direction within MM2. Indeed, based on the ALMA \ce{CS} 6-5 observation, \citet{2018ApJ...866...87B} found a one-sided blueshifted component from MM2 extending in the southwest direction. However, there is also a northeast–southwest outflow originating from MM1, and both the MM2 core and the one-sided outflow from MM2 are projected onto the large-scale blueshifted lobe of the MM1 outflow \citep{2018ApJ...866...87B}. Even though it is hard to constrain which one, or both, of the outflows is causing the shift of the AcA3 peaks, we can not rule out the role that $E_{\rm u}$ might have on the change in the AcA3 positions caused by outflows.

Although both of the mechanisms could account for the change in the AcA3 peak positions, the one that invokes dust opacity would lead to a correlation between positions and line frequencies, while the one that invokes outflows would lead to a correlation between positions and $E_\mathrm{u}$. With the current data, we are not able to constrain which one of the two is the dominant mechanism that causes the peak-position variations due to the limited number of clean lines and the coincidence that all the high $E_\mathrm{u}$ lines happen to be found at high frequencies. Nonetheless, the degeneracy could be resolved by imaging lines with low $E_\mathrm{u}$ and high frequencies, or vice versa. Moreover, our analysis is limited to the range of energy levels (154--324 K). Therefore, our data are probably not sufficient to exactly determine the dependence of peak positions on $E_{\rm u}$. Future work touching on precisely locating the emission region of more LAMs with a wider range of excitation energy is necessary, which could advance our knowledge to verify and understand this trend.

\section{SUMMARY}\label{sec:sum}
Based on ALMA observations of NGC 6334I, we have identified and imaged the first and second vibrationally excited states of \ce{CH3COOH} for the first time in the interstellar medium. More than 100 transitions attributed to vibrationally excited \ce{CH3COOH} contribute to the emission features in the 130-352 GHz spectra, although the majority of them are blended. It appears that vibrationally excited states of the detected interstellar molecules certainly account for a significant number of unidentified lines in the observed spectra toward hot core regions in molecular clouds. 

The observed spectrum was well-fit with the single excitation temperature model of \ce{CH3COOH} at the $v_\mathrm{t} = 1$ and $ v_\mathrm{t} = 2$ states. Through comparing the observed and modeled spectra, six features were assigned to the $v_\mathrm{t} = 1$ state while three features were assigned to the $v_\mathrm{t} = 2$ state. The best fit yields an excitation temperature of 142(25) K and a resulting column density of $1.12(7) \times 10^{17}~\mathrm{cm^{-2}}$.

We further imaged the peak intensity images of the 9 features that suffer least from contamination by other species. Through imaging multiple transitions, we eliminated spatial contamination in our analysis of the distribution of vibrationally excited \ce{CH3COOH}. Based on the intensity maps that show consistent morphology, we found that the \ce{CH3COOH} emissions are compact and concentrated toward the MM1 and MM2 regions. There is a global overlap of the \ce{CH3COOH} emission with the continuum emission, but, with a displacement between the molecular emission peaks and the continuum emission peaks. Three emission peaks of \ce{CH3COOH} were resolved. The AcA1 peaks are located toward the MM1A clump, and the AcA2 peaks lie between the MM1B and MM1D cores, while the AcA3 peaks are located toward the MM2 region. By locating the \ce{CH3COOH} emission from different values of $E_\mathrm{u}$, we found that there is not an obvious dependence of the positions of the AcA1 and AcA2 peak on excited energy levels, while the change of the AcA3 peak positions could be explained either by the effect of dust opacity or outflows. Our study does reveal the importance of the observation of a sufficient number of uncontaminated lines, which prevent our analysis from being biased by spatial contamination.

\acknowledgments
We acknowledge the support of the National Science Foundation (US) for the astrochemistry program of E. H. (grant number AST-1514844). This paper makes use of the following ALMA data: ADS/JAO.ALMA\#2015.A.00022.T, ADS/JAO.ALMA\#2017.1.00370.S and ADS/JAO.ALMA\#2017.1.00661.S. ALMA is a partnership of ESO (representing its member states), NSF (USA) and NINS (Japan), together with NRC (Canada), MOST and ASIAA (Taiwan), and KASI (Republic of Korea), in cooperation with the Republic of Chile. The Joint ALMA Observatory is operated by ESO, AUI/NRAO and NAOJ. The National Radio Astronomy Observatory is a facility of the National Science Foundation operated under cooperative agreement by the Associated Universities, Inc. Support for B.A.M. was provided by NASA through Hubble Fellowship grant \#HST-HF2-51396 awarded by the Space Telescope Science Institute, which is operated by the Association of Universities for Research in Astronomy, Inc., for NASA, under contract NAS5-26555. 

\software{CASA \citep{2007ASPC..376..127M}} 


\newpage

\appendix
\section{Spectroscopic Parameters of the Related Transitions\label{sec:transitions_detail}}
\begin{deluxetable}{clcc}[h!]
    \tablewidth{0pt}
    \tablehead{
        \\
        \colhead{Rest Frequency}    &\colhead{Quantum Number}   &\colhead{$E_u$}    &\colhead{$\log_{10}{\frac{A_{ul}}{\mathrm{s^{-1}}}}$}\\
        \colhead{(MHz)}             &~                          &\colhead{(K)}      &~                          
        }
    \startdata
	    130306.8457 &$v_\mathrm{t} = 1\ 11( 1,10)-10( 2, 9)\ A$  &154.261    &-4.52\\
	    130306.8458 &$v_\mathrm{t} = 1\ 11( 2,10)-10( 2, 9)\ A$  &"          &-5.82\\
        130306.8462 &$v_\mathrm{t} = 1\ 11( 1,10)-10( 1, 9)\ A$  &"          &-5.82\\
        130306.8462 &$v_\mathrm{t} = 1\ 11( 2,10)-10( 1, 9)\ A$  &"          &-4.52\\
        144791.7911 &$v_\mathrm{t} = 2\ 13( 0,13)-12( 1,12)\ E$  &246.132    &-4.41\\
        144791.8037 &$v_\mathrm{t} = 2\ 13( 1,13)-12( 1,12)\ E$  &"          &-5.19\\
        144791.8596 &$v_\mathrm{t} = 2\ 13( 0,13)-12( 0,12)\ E$  &"          &-5.19\\
        144791.8722 &$v_\mathrm{t} = 2\ 13( 1,13)-12( 0,12)\ E$  &"          &-4.41\\
        144975.0345 &$v_\mathrm{t} = 2\ 13( 0,13)-12( 1,12)\ A$  &226.084    &-4.48\\
        144975.0422 &$v_\mathrm{t} = 2\ 13( 1,13)-12( 1,12)\ A$  &"          &-4.82\\
        144975.0594 &$v_\mathrm{t} = 2\ 13( 0,13)-12( 0,12)\ A$  &"          &-4.82\\
        144975.0672 &$v_\mathrm{t} = 2\ 13( 1,13)-12( 0,12)\ A$  &"          &-4.48\\
        251599.6383 &$v_\mathrm{t} = 2\ 23( 0,23)-22( 0,22)\ A$  &323.807    &-3.77\\
        "           &$v_\mathrm{t} = 2\ 23( 1,23)-22( 1,22)\ A$  &"          &-3.77\\
        "           &$v_\mathrm{t} = 2\ 23( 0,23)-22( 1,22)\ A$  &"          &-4.04\\
        "           &$v_\mathrm{t} = 2\ 23( 1,23)-22( 0,22)\ A$  &"          &-4.04\\
        266285.3833 &$v_\mathrm{t} = 1\ 22( 3,19)-21( 4,18)\ A$  &274.159    &-3.79\\
        "           &$v_\mathrm{t} = 1\ 22( 4,19)-21( 3,18)\ A$  &"          &-3.79\\
        "           &$v_\mathrm{t} = 1\ 22( 3,19)-21( 3,18)\ A$  &"          &-3.99\\
        "           &$v_\mathrm{t} = 1\ 22( 4,19)-21( 4,18)\ A$  &"          &-3.99\\
        270849.2061 &$v_\mathrm{t} = 1\ 22( 3,19)-21( 4,18)\ E$  &270.833    &-3.71\\
        "           &$v_\mathrm{t} = 1\ 22( 4,19)-21( 3,18)\ E$  &"          &-3.71\\
        "           &$v_\mathrm{t} = 1\ 22( 3,19)-21( 3,18)\ E$  &"          &-4.09\\
        "           &$v_\mathrm{t} = 1\ 22( 4,19)-21( 4,18)\ E$  &"          &-4.09\\
        281447.5805 &$v_\mathrm{t} = 1\ 23( 3,20)-22( 4,19)\ E$  &284.341    &-3.63\\
        "           &$v_\mathrm{t} = 1\ 23( 4,20)-22( 3,19)\ E$  &"          &-3.63\\
        "           &$v_\mathrm{t} = 1\ 23( 3,20)-22( 3,19)\ E$  &"          &-4.10\\
        "           &$v_\mathrm{t} = 1\ 23( 4,20)-22( 4,19)\ E$  &"          &-4.10\\
        281789.7257 &$v_\mathrm{t} = 1\ 24( 2,22)-23( 2,21)\ E$  &287.659    &-3.59\\
        "           &$v_\mathrm{t} = 1\ 24( 3,22)-23( 3,21)\ E$  &"          &-3.59\\
        "           &$v_\mathrm{t} = 1\ 24( 2,22)-23( 3,21)\ E$  &"          &-4.11\\
        "           &$v_\mathrm{t} = 1\ 24( 3,22)-23( 2,21)\ E$  &"          &-4.11\\
        282044.2002 &$v_\mathrm{t} = 1\ 25( 1,24)-24( 2,23)\ E$  &289.904    &-3.61\\
        "           &$v_\mathrm{t} = 1\ 25( 2,24)-24( 1,23)\ E$  &"          &-3.61\\
        "           &$v_\mathrm{t} = 1\ 25( 1,24)-24( 1,23)\ E$  &"          &-3.97\\
        "           &$v_\mathrm{t} = 1\ 25( 2,24)-24( 2,23)\ E$  &"          &-3.97
    \enddata
    \tablecomments{Relevant spectroscopic data of the transitions corresponding to the nine clean emission features in Table \ref{v-table} taken from \citet{2013JMoSp.290...31I}.}
\end{deluxetable}

\newpage

\section{The Whole Spectra \label{sec:a-spect}}
\begin{figure*}[htp!]
\epsscale{1.2}
\plotone{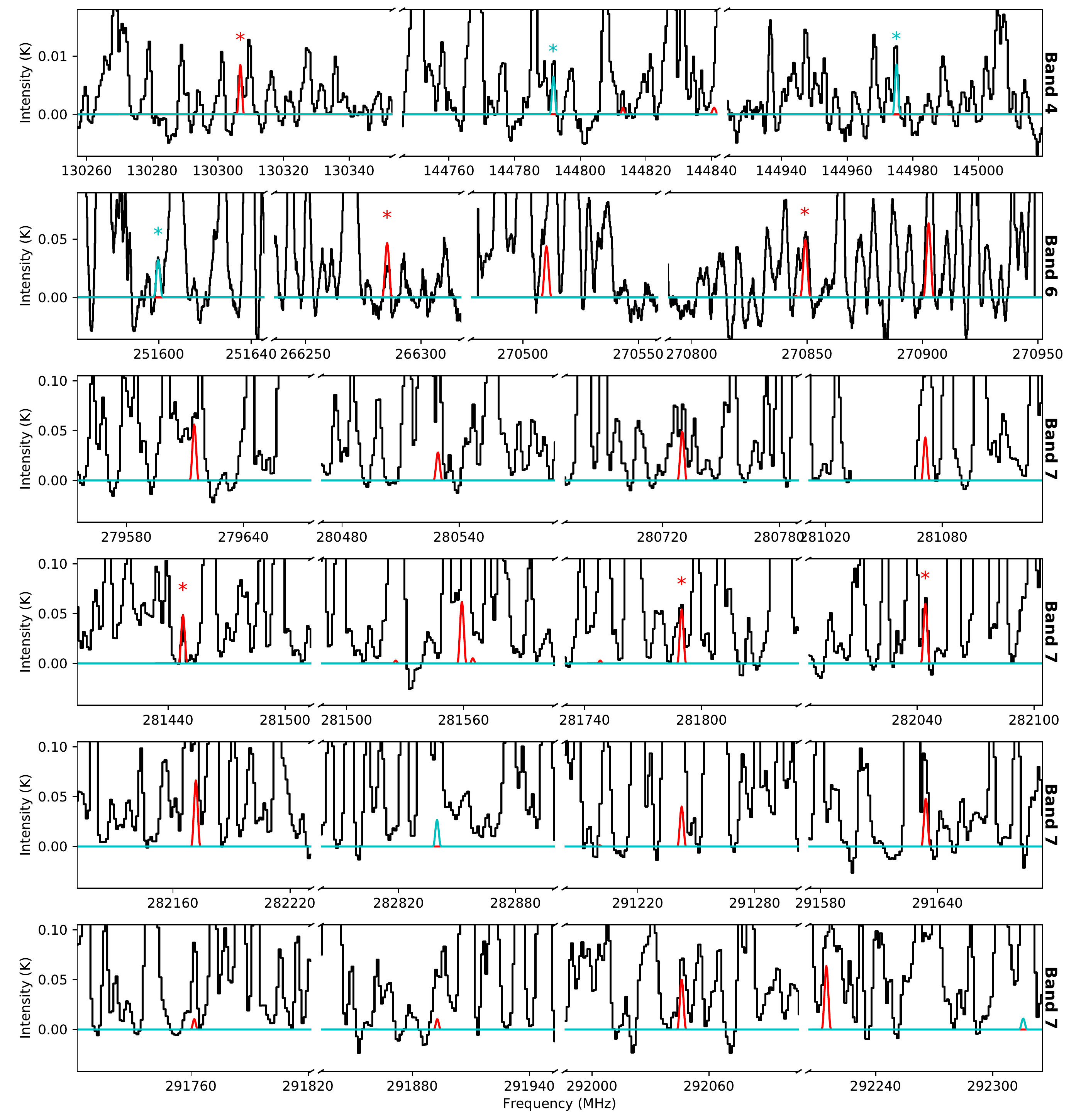}
\caption{The observed spectra of NGC 6334I extracted toward the MM1 region are plotted in black with the single-excitation temperature model spectra of vibrationally excited \ce{CH3COOH} in red for $v_\mathrm{t} = 1$ and cyan for $v_\mathrm{t} = 2$. Asterisk makers mark the clean spectral features, of which the blowups are shown in Figure \ref{v1_and_2-spec}. The best-fit parameters are $T_\mathrm{ex} = 142~\mathrm{K}$, $N_\mathrm{T} = 1.12 \times 10^{17}~\mathrm{cm^{-2}}$, and $\Delta V = 2.72~\mathrm{km\,s^{-1}}$. The horizontal axes are the rest frequency with respect to a radial velocity of $-6.9~\mathrm{km\,s^{-1}}$.
\label{v1_and_2-spec-1}}
\end{figure*}

\begin{figure}
\epsscale{1.2}
\plotone{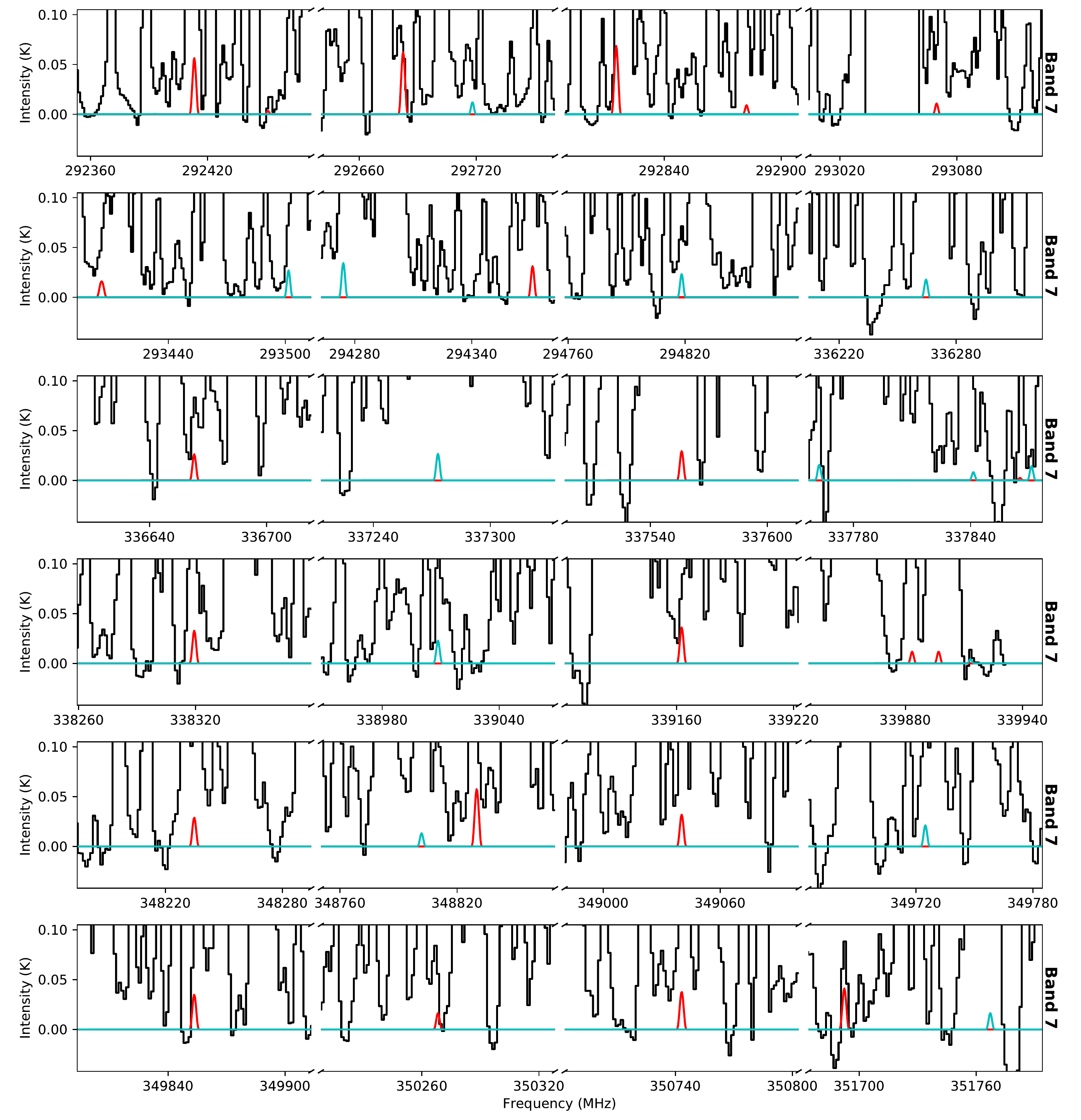}
\caption{Continue to Figure \ref{v1_and_2-spec-1}\label{v1_and_2-spec-2}}
\end{figure}

\end{document}